\documentclass[12pt]{article}
\newcommand{\blind}{0} %Set this to 1 for blinded version
\usepackage{times}
\usepackage{epsfig}
\usepackage{float}
\usepackage{amsfonts}
\usepackage{bm,pbox}
\usepackage{multirow} 
\usepackage{color}
\usepackage{framed}
\usepackage{amssymb,amsthm,amsfonts,amsbsy,latexsym,dsfont}
\usepackage{natbib}
\usepackage[usenames,dvipsnames,svgnames,table]{xcolor}

\usepackage{hyperref}
\hypersetup{colorlinks,breaklinks,
            linkcolor=blue,urlcolor=blue,
            anchorcolor=darkblue,citecolor=blue}

\usepackage{amssymb}
\usepackage{amsmath}
\usepackage{enumerate}
\usepackage{epstopdf}
\usepackage{graphicx}
\usepackage{caption}
\usepackage{subcaption}
\usepackage{verbatim}
\usepackage{MJDArticle}
\usepackage[margin=1in]{geometry}
\newcommand{\mvLambda}{\boldsymbol{\beta}}

\newtheorem{@assumption}{ Assumption}[section] 

\usepackage{fixltx2e}

\DeclareMathOperator{\argmax}{argmax}

\usepackage{bm}% bold math
\newcommand{\bt}{\pmb{\theta}}

\newcommand{\bh}{\pmb{\text{h}}}

\def\spacingset#1{\renewcommand{\baselinestretch}%
{#1}\small\normalsize} \spacingset{1}

\newcommand{\set}[1]{\left\{#1\right\}}

\usepackage{todonotes}
\usepackage{setspace}
\usepackage{framed}
\usepackage{soul}
\setstcolor{red}
\usepackage{color}

\begin{document}
\if0\blind
{
  \if0\blind
  {
    \title{\bf Stochastic Weighted Graphs: Flexible Model Specification and Simulation\thanks{
      The authors gratefully acknowledge the National Science Foundation and NSF grants DMS-1105581, DMS-1310002, SES-1357622, SES-1357606, SES-1461493, and CISE-1320219.}} 
   
    \author{James D. Wilson, Matthew J. Denny, Shankar Bhamidi,\\
    Skyler Cranmer, and Bruce Desmarais\footnote{James D. Wilson is an Assistant Professor in the Department of Mathematics and Statistics at the University of San Francisco (\href{jdwilson4@usfca.edu}{jdwilson4@usfca.edu}). Matthew Denny is a Ph.D. Candidate in the Department of Political Science and Social Data Analytics at Pennsylvania State University (\href{mdenny@psu.edu}{mdenny@psu.edu}). Shankar Bhamidi is an Assistant Professor in the Department of Statistics and Operations Research at the University of North Carolina (\href{bhamidi@email.unc.edu}{bhamidi@email.unc.edu}). Skyler Cranmer is the Carter Phillips and Sue Henry Associate Professor in the Department of Political Science at the Ohio State University (\href{cranmer.12@osu.edu}{cranmer.12@osu.edu}). Bruce Desmarais is an Associate Professor in the Department of Political Science at Pennsylvania State University (\href{bbd5087@psu.edu}{bbd5087@psu.edu}).}}
 % \author{\footnotesize James D. Wilson\thanks{
  %  The authors gratefully acknowledge the National Science Foundation and NSF grants DMS-1105581, DMS-1310002, SES-1357622, SES-1357606, and CISE-1320219.} \hspace{.2cm}\\
   % {\small Department of Statistics and Operations Research, University of North Carolina}\\
   % Matthew J. Denny \\
   % {\small Department of Political Science, University of Massachusetts, Amherst}\\
%	Shankar Bhamidi\\
%	{\small Department of Statistics and Operations Research, University of North Carolina}\\
%	Skyler Cranmer\\
%	{\small Department of Political Science, The Ohio State University}\\
%	and\\
%	Bruce Desmarais\\
%	{\small Department of Political Science, University of Massachusetts, Amherst}}
  \maketitle
} \fi

\if1\blind
{
  \bigskip
  \bigskip
  \bigskip
  \begin{center}
    {\LARGE\bf Title}
\end{center}
  \medskip
} \fi

\vskip 1pc
\begin{abstract}
	
\noindent In most domains of network analysis researchers consider networks that arise in nature with weighted edges. Such networks are routinely dichotomized in the interest of using available methods for statistical inference with networks. The generalized exponential random graph model (GERGM) is a recently proposed method used to simulate and model the edges of a weighted graph. The GERGM specifies a joint distribution for an exponential family of graphs with continuous-valued edge weights. However, current estimation algorithms for the GERGM only allow inference on a restricted family of model specifications. To address this issue, we develop a Metropolis--Hastings method that can be used to estimate any GERGM specification, thereby significantly extending the family of weighted graphs that can be modeled with the GERGM. We show that new flexible model specifications are capable of avoiding likelihood degeneracy and efficiently capturing network structure in applications where such models were not previously available. We demonstrate the utility of this new class of GERGMs through application to two real network data sets, and we further assess the effectiveness of our proposed methodology by simulating non-degenerate model specifications from the well-studied two-stars model.  A working R version of the GERGM code is available in the supplement and will be incorporated in the {\it gergm} CRAN package. 
	
\end{abstract}

\noindent%
{\it Keywords:} Exponential Random Graph, Generalized Exponential Random Graph, Markov Chain Monte Carlo, Metropolis--Hastings
\vfill

\newpage
\spacingset{1.45} % DON'T change the spacing!

%%%%%%%
%%Old title
%\title{\bf Complex Stochastic Weighted Graphs:\\ Flexible Model Specification and Simulation%\footnote{AMS subject classification. Primary: ***. Secondary: ***.}
%\footnote{James D. Wilson is a Ph.D. Candidate in the Department of Statistics and Operations Research at the University of North Carolina at Chapel Hill (\href{jameswd@email.unc.edu}{jameswd@email.unc.edu}). Matthew Denny is a Ph.D. Candidate in the Department of Political Science at the University of Massachusetts, Amherst (\href{mdenny@polsci.umass.edu}{mdenny@polsci.umass.edu}). Shankar Bhamidi is an Assistant Professor in the Department of Statistics and Operations Research at the University of North Carolina (\href{bhamidi@email.unc.edu}{bhamidi@email.unc.edu}). Skyler Cranmer is the Carter Phillips and Sue Henry Associate Professor in the Department of Political Science at the Ohio State University (\href{cranmer.12@osu.edu}{cranmer.12@osu.edu}). Bruce Desmarais is an Assistant Professor in the Department of Political Science at the University of Massachusetts, Amherst (\href{desmarais@polsci.umass.edu}{desmarais@polsci.umass.edu}).} 
%\footnote{All code used in this manuscript is available in the R package {\bf xergm}. Code to reproduce all results are given in the supplemental material.}\footnote{{\it Keywords}: exponential random graph model, generalized exponential random graph model,  markov chain monte carlo, weighted graphs.}}

%\author{James D. WILSON, Matthew J. DENNY, Shankar BHAMIDI,\\
%Skyler CRANMER, and Bruce DESMARAIS}
%\affil{University of North Carolina at Chapel Hill}
%\affil{University of Massachusetts Amherst}
%\affil{The Ohio State University}
%\maketitle
%%%%%%
\section{Introduction}\label{sec:intro}
% Networks are used in a wide variety of fields such as sociology, genetics, finance, and neuroscience \citep[see, e.g., ][, respectively]{wasserman1994social,Zhang:2005,Iori:2008,Greicius:2003}. The analysis of network data, however, presents unique challenges in modeling and simulation that require innovative statistical and computational techniques (see \cite{goldenberg2010survey, hunter2012computational, fienberg2012brief} for recent discussions).

Throughout the sciences, but particularly in the social sciences, a fundamental tool for the statistical analysis of networks has been the exponential random graph model (ERGM) - a popular, powerful, and flexible tool for statistical inference with network data \citep{holland1981exponential, Wasserman:1996, Snijders:2006}. Despite their popularity, conventionally used ERGMs have the major limitation that they require the edges of an observed network be binary (representing the presence or absence of an edge). Thus ERGMs cannot directly model weighted networks. Since many substantively important networks are weighted, this restriction is especially problematic. Weighted networks arise, for example, in the study of financial exchange \citep{Iori:2008}, migration patterns \citep{Chun:2008}, and in the analysis of brain functionality and connectivity \citep{Simpson:2011}. Recently, some progress on modeling weighted networks in the ERGM framework was made in \cite{Desmarais:2012}, where the generalized exponential random graph model (GERGM) was proposed to study networks with continuous-valued edges. Around the same time, \cite{krivitsky2012exponential} proposed the weighted exponential random graph model that generalized the ERGM to networks with integer-valued edges. \cite{robins1999logit} developed logistic dyad-independent models for networks with integer-valued edges. Though each of these models provide a means to analyze weighted networks, we will focus on extensions to the GERGM.

In general, the likelihood function of an ERGM is intractable (though some recent progress has been made in the large network $n\to\infty$ limit \citep{chatterjee2013estimating,lubetzky2014replica}); however, efficient estimation can be achieved through the use of Markov Chain Monte Carlo (MCMC) algorithms \citep{geyer1992constrained, Hunter:2006}.  MCMC can be used to simulate samples of networks from which the likelihood function of an ERGM can be approximated. Like the ERGM, estimation of the GERGM is readily achieved via MCMC algorithms. \cite{Desmarais:2012} proposed a Gibbs sampling technique for GERGM estimation; however, this strategy limits the specification of network dependencies captured by the GERGM to those for which full conditional edge distributions can be derived in closed form. Another important obstacle that arises in discrete exponential family model specification is the problem of degeneracy, a condition under which only a few network configurations - usually very sparse and very dense networks - have high probability mass \citep{handcock2003assessing, rinaldo2009geometry, schweinberger2011instability}. The issue of degeneracy strongly influences the effectiveness of an MCMC algorithm. Indeed, in the case that nearly empty (or nearly complete) networks are most probable, estimation via MCMC will fail to converge to consistent parameter estimates.

Here, we expand the family of weighted networks that can be analyzed under the GERGM by developing a Metropolis--Hastings sampling procedure that allows the flexible specification of network statistics and models under the GERGM framework. Perhaps the greatest drawback of the limited set of models for which Gibbs Sampling can be used to simulate networks is that they are prone to degeneracy. This is due to the fact that the closed-form derivation of the conditional distribution of an edge requires that the network statistics used to specify the GERGM depend linearly on the value of each edge. GERGM specifications that include nonlinear statistics are often required to avoid degeneracy.  A significant advantage of our proposed Metropolis--Hastings (MH) procedure is that one can use MH sampling to estimate models that involve nonlinear network statistics. The expanded set of GERGM specifications made available with the use of MH can be used to find a non-degenerate model specification. Furthermore, in models where the Gibbs sampler can be used, Metropolis--Hastings yields the same parameter estimates as those obtained via Gibbs. The framework established here provides an important step in flexibly modeling and simulating weighted networks while further providing a means of avoiding model degeneracy.

In Section \ref{sec:GERGM}, we describe the generalized exponential random graph model for graphs with continuous-valued edges. In Section \ref{sec:inference}, we discuss the Monte Carlo maximum likelihood estimation of the GERGM and briefly describe the Gibbs procedure devised in \cite{Desmarais:2012}. At the end of Section \ref{sec:inference}, we formulate a flexible Metropolis--Hastings sampling procedure. We propose a class of model specifications in Section \ref{sec:networkstats} that expands the family of GERGMs beyond those permissible under Gibbs sampling. In Section \ref{sec:numerical}, we evaluate the performance and potential utility of our proposed framework through application to the U.S. state-to-state migration network, an international financial exchange network, as well as through a simulation study that revisits the degenerate two-star-model of \cite{handcock2003assessing}. We conclude with a discussion of open problems and future work in Section \ref{sec:discussion}.

%%%Section: The Generalized Exponential Random Graph Model
\section{The Generalized Exponential Random Graph Model}\label{sec:GERGM} 

Consider a directed network defined on a node set $[n]=\set{1,2,\ldots,n}$, where $m = n(n-1)$ denotes the total number of directed edges between these nodes. Suppose that the weighted relationships between the nodes are represented by a collection of weights $(y_{ij}:i\neq j\in [n])\in \mathbb{R}^m$. The aim of this section is to describe a specific class of probability models on $\mathbb{R}^{m}$ as constructed in \cite{Desmarais:2012} called GERGMs that incorporates relational structure between the nodes to generate a random vector $Y \in \mathbb{R}^m$. This probability distribution is specified by a joint probability density function (pdf) $f_Y(y , \Theta)$ driven by real-valued parameters $\Theta$. 

A GERGM for the observed configuration $y$ has a simple generative process that relies on two distinct steps. First, a joint distribution that captures the structure and interdependence of $Y$ is defined on a restricted network configuration, $X \in [0,1]^m$. Next, the restricted network $X$ is transformed onto the support of $Y$ through an appropriate transformation function. These two steps are closely related to the widely studied specification of joint distributions via copula functions \citep{Genest1986}. We now describe the two steps in specifying a GERGM in more detail.

In the first specification step, a function of network summary statistics $\bh: [0,1]^m \rightarrow \mathbb{R}^p$ is formulated to represent the joint features of $X$. The random vector $X$ is modeled by an exponential family with parameters $\bt \in \mathbb{R}^p$ as follows:

%f_X%
\begin{equation}
 f_X(x,\bt) = \frac{\exp \left(\bt' \bh(x)\right)}{\int_{[0,1]^m} \exp\left(\bt' \bh(z)\right)dz},  \text{\hskip 1pc $x \in [0,1]^m$},
 \label{eq:fofx} 
\end{equation} 
\noindent where $\bm{\theta}'$ denotes the transpose of the vector $\bm{\theta}$. The network specification in model (\ref{eq:fofx}) is closely related to the usual specification of exponential random graph models on binary edges with the exception that individual edges are now modeled as having continuous weights taking values between 0 and 1. As dependence relationships can be captured by functions of edges valued on the unit interval, model (\ref{eq:fofx}) provides a flexible specification of interdependence. For instance, networks generated by a highly reciprocal process are likely to exhibit high values of $\sum_{i < j}x_{ij}x_{ji}$, and those for which there is a high variance in the popularity of vertices (e.g., preferential attachment) are likely to exhibit high values of the ``two-stars" statistic $\sum_{i}\sum_{j,k \neq i}x_{ji}x_{ki}$ \citep{Park:2004b}. We describe several flexible network statistics for modeling interdependence in Section \ref{sec:networkstats}. % An important feature of (\ref{eq:fofx}) is that when the statistics $\bh(X)$ play no role in the structure of $X$, then the edges $\{X_{ij}\}$ are modeled as independent uniform random variables on the unit interval.
Note that the uniform distribution on $[0,1]^m$ is a special case of the model above, obtained by setting the parameters $\bt = \bm{0}$.

In the second specification step, a one-to-one and coordinate wise monotonically non-decreasing function $T^{-1}: [0,1]^m \rightarrow \mathbb{R}^m$ is formulated to model the transformation of the restricted network $X$ onto the support of $Y$. Specifically, for each pair of distinct nodes $i$, $j \in [n]$, we model $Y_{ij} = T^{-1}_{ij}(X, \bm{\beta})$ where $\bm{\beta} \in \mathbb{R}^{k}$ parameterizes the transformation so as to capture the marginal features of $Y$. Since $T^{-1}$ is a monotonically non-decreasing, the pdf of $Y$ is given by

%f_Y( )%
\begin{equation}
f_Y(y ,\bt, \bm{\beta}) = \frac{\exp \left(\bt' \bh(T(y,\bm{\beta}))\right)}{\int_{[0,1]^m} \exp\left(\bt' \bh(z)\right)dz}\prod_{ij}t_{ij}(y,\bm{\beta}), \text{\hskip 1pc $y \in \mathbb{R}^m$} 
\label{eq:fofy} 
\end{equation} 

\noindent where $t_{ij}(y, \bm{\beta}) = dT_{ij}(y,\bm{\beta})/dy_{ij}$. Though the choice of $T^{-1}$ is flexible, specifying $T^{-1}$ so that $T^{-1}_{ij}$ is an inverse cumulative distribution function (cdf) is advisable because the properties of \eqref{eq:fofy} are difficult to understand without this restriction and because it leads to several beneficial properties. First, when $T^{-1}$ is an inverse cdf, $t_{ij}$ is precisely a marginal pdf for all $i \neq j$. Second, when $\bt = \bm{0}$, then $f_Y(y, \bt, \bm{\beta})$ reduces to a product of marginal pdfs $\{t_{ij}\}$ and thus in this special case one obtains a model with dyadic independence acorss edge weight distributions. An important example includes taking $T^{-1}$ as the inverse of a Gaussian cdf with constant variance. In this special case, if $\bt = \bm{0}$ then (\ref{eq:fofy}) reduces to a model for conditionally independent Gaussian observations, such as ordinary least squares regression.

%%Subsection: Maximum Likelihood Estimation
\section{Model Inference}\label{sec:inference}
The GERGM specification in equations (\ref{eq:fofx}) and (\ref{eq:fofy}) can be used to readily model a wide range of network interdependencies in weighted networks. In this section, we describe maximum likelihood inference of the parameters $\bt$ and $\bm{\beta}$ via MCMC. We review the Gibbs sampling procedure in \cite{Desmarais:2012}, which relies on an important restriction of model specification. We then develop a general inferential framework for sampling via Metropolis--Hastings, which extends the family of GERGM specifications. We provide pseudo-code for the MCMC maximum likelihood estimation procedure described in Sections 3.1 - 3.4 in the Appendix. 

\subsection{Maximum Likelihood Inference}

Given a specification of statistics $\bh(\cdot)$, transformation function $T^{-1}$, and observations $Y=y$ from the distribution (\ref{eq:fofy}), our goal is to find the maximum likelihood estimates (MLEs) of the unknown parameters $\bt$ and $\bm{\beta}$, namely to find values $\widehat{\bt}$ and $\widehat{\bm{\beta}}$ that maximize the log likelihood:
%Log Likelihood
\begin{equation}\label{eq:loglike}
\ell(\bt, \bm{\beta} | y) = \bt'\bh(T(y,\bm{\beta})) - \log C(\bt) + \sum_{ij} \log t_{ij}(y, \bm{\beta}),
\end{equation}
where
$$C(\bt) = \int_{[0,1]^m} \exp(\bt' \bh(z))dz.$$

The maximization of (\ref{eq:loglike}) can be achieved through alternate maximization of $\bm{\beta} | \bt$ and $\bt | \bm{\beta}$. In particular, one can calculate the MLEs $\widehat{\bt}$ and $\widehat{\bm{\beta}}$ by iterating between the following two steps until convergence.

%MLE algorithm 
\noindent For $r \geq 1$, iterate until convergence:
\begin{enumerate}
\item Given $\bt^{(r)}$, estimate $\bm{\beta}^{(r)} | \bt^{(r)}$:
%Lambda optimization
\begin{equation}\label{eq:lambda}
\mvLambda^{(r)} = \argmax_{\bm{\beta}} \left(\bt^{(r)}\bh(T(y,\bm{\beta})) + \sum_{ij} \log t_{ij}(y,\bm{\beta})\right).
\end{equation}
%Theta optimization
\item Set $\hat{x} = T(y,{\bm{\beta}}^{(r)})$. Then estimate $\bt^{(r+1)} | \bm{\beta}^{(r)}$:
\begin{equation}\label{eq:theta}
{\bt}^{(r+1)} = \argmax_{\bt} \Big( \bt'\bh(\hat{x}) - \log C(\bt) \Big).
\end{equation}
\end{enumerate}

For fixed $\bt$, the likelihood maximization in (\ref{eq:lambda}) is straightforward and can be accomplished numerically using gradient descent \citep{snyman2005practical}. In the case that $t_{ij}$ is log-concave and $\bh \circ T$ is concave in $\bm{\beta}$, a hill climbing algorithm is assured to find the global optimum. 

The maximization in (\ref{eq:theta}) is a difficult problem due to the intractability of the normalization factor $C(\bt)$. There has been much recent work on circumventing the intractability of $C(\bt)$. For example, \citet{strauss1990pseudolikelihood} consider using the maximim pseudo-likelihood estimate (MPLE) for $\bt$, which assumes independence of the edges in the graph. \citet{van2009framework} shows, however, that using the MPLE is often biased and far less efficient than the maximum likelihood estimate especially when strong network dependencies are present. In light of the inefficiency of pseudo-likelihood estimates, we turn to MCMC methods for estimating (\ref{eq:theta}) which have witnessed considerable success in estimating exponential family models \citep{geyer1992constrained, Hunter:2006}. We describe the  
MCMC framework for estimating $\bt$ and then review the constrained Gibbs procedure developed in \citet{Desmarais:2012} before introducing our new more flexible Metropolis--Hastings procedure.

%%Subsection: Monte Carlo Maximization
\subsection{Monte Carlo Maximization in the GERGM}
Let $\bt$ and $\widetilde{\bt}$ be two arbitrary vectors in $\mathbb{R}^p$ and let $C(\cdot)$ be defined as in (\ref{eq:loglike}). The crux of optimizing (\ref{eq:theta}) via Monte Carlo simulation relies on the following property of exponential families \citep{geyer1992constrained}:

%%Geyer and Thompson fact
\begin{equation} \label{eq:ratio_odds}
\dfrac{C({\bt})}{C(\widetilde{\bt})} = \mathbb{E}_{\widetilde{\bt}} \left[\exp\left((\bt - \widetilde{\bt})'  \bh(X) \right)\right].
\end{equation}

\noindent The expectation in (\ref{eq:ratio_odds}) is not directly computable; however, a first order approximation to this quantity is given by the first moment estimate:

%%First moment approximation
\begin{equation} \label{eq:approximation}
\mathbb{E}_{\widetilde{\bt}} \left[\exp\left((\bt - \widetilde{\bt})'  \bh(X) \right)\right] \approx \dfrac{1}{M} \sum_{j = 1}^M \exp\left( (\bt - \widetilde{\bt})'  \bh(x^{(j)}) \right), \end{equation}
where $x^{(1)}, \ldots, x^{(M)}$ is an observed sample from pdf $f_X(\cdot, \widetilde{\bt})$. 

Define $\ell(\bt | \hat{x}): = \bt \bh(\hat{x}) - \log C(\bt)$. Then maximizing $\ell(\bt | \hat{x})$ with respect to $\bt \in \mathbb{R}^p$ is equivalent to maximizing $\ell(\bt | \hat{x}) - \ell(\widetilde{\bt} | \hat{x})$ for any fixed arbitrary vector $\widetilde{\bt} \in \mathbb{R}^p$. Equations (\ref{eq:ratio_odds}) and (\ref{eq:approximation}) suggest:

%%Approximation to the difference in likelihoods
\begin{equation} \label{eq:log_diff}
\ell(\bt | \hat{x}) - \ell(\widetilde{\bt} | \hat{x}) \approx (\bt - \widetilde{\bt})' \bh(\hat{x}) - \log \left( \dfrac{1}{M} \sum_{j = 1}^M \exp\left( (\bt - \widetilde{\bt})'  \bh(x^{(j)}) \right) \right).
\end{equation}

\noindent An estimate for $\bt$ can now be calculated by the maximization of (\ref{eq:log_diff}). %Consider the $r$th iterate update of $\bt$ from (\ref{eq:theta}). 
The $r+1$st iteration estimate $\bt^{(r+1)}$ in \ref{eq:theta} can be obtained using Monte Carlo methods by iterating between the following two steps:

%MCMCMLE Algorithm
\noindent Given $\bm{\beta}^{(r)}$, $\bt^{(r)}$, and $\hat{x} = T(y,\bm{\beta}^{(r)})$
\begin{enumerate}
	\item Simulate networks $x^{(1)}, \ldots, x^{(M)}$ from density $f_X(x~,~\bt^{(r)}).$
	\item Update:
	\begin{equation}\label{eq:montecarlo}\bt^{(r+1)} = \argmax_{\bt} \left(\bt' \bh(\hat{x}) - \log \left( \dfrac{1}{M} \displaystyle\sum_{j = 1}^M \exp\left( (\bt - \bt^{(r)})'  \bh(x^{(j)}) \right) \right)\right).\end{equation}
	\end{enumerate}
%End Algorithm

%initialization
%\todo[inline]{IMPORTANT, IMPORTANT, IMPORTANT: If we have observations $Y=y$ from general $\mathbb{R}^m$, how are we getting our initial $x_{obs}$? What is the initialization for $\beta_0$ so that we can plug in $x_{obs} = T(y,\beta_0)$? Also provide some text from the Snijders book as to why this is a reasonable thing to do.}

Given observations $Y = y$, the Monte Carlo algorithm described above requires an initial estimate $\bm{\beta}^{(0)}$ and $\bt^{(1)}$. We initialize $\beta_0$ using (\ref{eq:lambda}) in the case that there are no network dependencies present, namely, $\bm{\beta}^{(0)} = \argmax_{\bm{\beta}}\{\sum_{ij} \log t_{ij}(y, \bm{\beta})\}$. We then fix $x_{obs} = T(y, \beta_0)$, and use the Robbins-Monro algorithm for exponential random graph models described in \cite{snijders2002markov} to initialize $\bt^{(1)}$. This initialization step can be thought of as the first step of a Newton-Raphson update of the MPLE estimate $\bt_{MPLE}$ on a small sample of networks generated from the density $f_X(x_{obs}, \bt_{MPLE})$.

%JW: I've decided to take this out as I think that it makes the discussion hard to follow. 

% The initialization procedure can be described as follows. First the MPLE, $\widehat{\bt}$, is calculated using $x_{obs}$. Then a sample of $M_1$ networks, $x^{(1)}, \ldots, x^{(M_1)}$, is simulated from $f_{X}(x_{obs}, \hat{\bt})$. The expected vector of statistics $\bar{\bh}$ and the empirical covariance matrix $D$ of the network sample is then calculated.
% $$\bar{\bh} := M_1^{-1} \sum_{i = 1}^{M_1} \bh(x^{(i)})$$
% $$D := M_1^{-1} \sum_{i = 1}^{M_1} \bh(x^{(i)}) \bh(x^{(i)})' - \bar{\bh}\bar{\bh}'$$
%
% \noindent According to the Robbins-Monro algorithm, $\bt^{(1)}$ is then calculated through the use of a modified Newton-Raphson update that compares the network samples to the observed network:
%
% \begin{equation} \bt^{(0)} = \widehat{\bt} - a D^{-1} (\bar{\bh} - \bh(x_{obs})) \end{equation}
%
% \noindent where $a \in [0,1]$ is the gain factor that controls the extremity of the distance between $\bt^{(0)}$ and the MPLE $\widehat{\bt}$. By default, we follow the suggestion of \cite{lusher2012exponential} and use $a = 0.10$.

The first step of the Monte Carlo algorithm requires simulation from the density $f_X({x}, \bt^{(r)})$. As this density cannot be directly computed, one must rely on the use of MCMC methods, such as Gibbs or Metropolis--Hastings samplers, for estimation.

%%Subsection: Gibbs Sampler
\subsection{Simulation via Gibbs Sampling}
The Gibbs sampling procedure described in \cite{Desmarais:2012} provides a straightforward way to estimate $\bt$ through the iterative optimization of (\ref{eq:log_diff}); however, its use restricts the specification of network statistics $\bh(\cdot)$ in the GERGM formulation. In particular, the use of Gibbs sampling requires that the network dependencies in an observed network $y$ are captured through $x$ by first order network statistics, namely statistics $\bh(\cdot)$ that are linear in $x_{ij}$ for all $i,j \in [n]$. With this assumption, one can derive a closed-form conditional distribution of $X_{ij}$ given the remaining network, $X_{-(ij)}$, which is used in Gibbs sampling.

Let $f_{X_{ij} | X_{-(ij)}}(x_{ij}, \bt)$ denote the conditional pdf of $X_{ij}$ given the remaining restricted network $X_{-(ij)}$. Consider the following condition on $\bh(x)$:

\begin{equation}\label{eq:assumption} \frac{\partial^2 \bh(x)}{\partial x^2_{ij}} = \bm{0}, \text{\hskip 1pc $i,j \in [n]$} \end{equation}

\noindent Assuming that (\ref{eq:assumption}) holds, one can readily derive a closed form expression for $f_{X_{ij} | X_{-(ij)}}(x_{ij}, \bt)$:

%%Conditional density
\begin{equation}\label{eq:cond_density} f_{X_{ij} | X_{-(ij)}}(x_{ij}, \bt) =   \frac{\exp\left(x_{ij} \frac{\bt' \partial \bh(x)}{\partial x_{ij}}\right)}{\left(\bt'\frac{\partial \bh(x)}{\partial x_{ij}}\right)^{-1}\left[ \exp(\bt' \frac{\partial \bh(x)}{\partial x_{ij}})-1\right]} \end{equation}

Let $U$ be uniform on $(0,1)$. Using the conditional density in (\ref{eq:cond_density}), one can simulate values of $x \in \mathbb{R}^m$ iteratively by drawing edge realizations of $X_{ij} | X_{-(ij)}$ according to the following distribution:

\begin{equation}\label{eq:cond_sim} X_{ij} | X_{-ij} \sim \dfrac{\log \left[1 + U \left(\exp(\bt' \frac{\partial \bh(x)}{\partial x_{ij}}) - 1\right)\right]}{\bt' \frac{\partial \bh(x)}{\partial x_{ij}}}, \text{\hskip 1pc $\bt' \frac{\partial \bh(x)}{\partial x_{ij}} \neq 0$} \end{equation}
When $\bt' \frac{\partial \bh(x)}{\partial x_{ij}} = 0$, all values in [0,1] are equally likely; thus, $X_{ij}|X_{-(ij)}$ is simply drawn uniformly from support [0,1]. The Gibbs simulation procedure simulates network samples $x^{(1)}, \ldots, x^{(M)}$ from $f_X(x,\bt)$ by sequentially sampling each edge from its conditional distribution given in  (\ref{eq:cond_sim}).

%\todo[inline]{Please feel free to modify or strengthen the following paragraph. I think that statistical interactions and degeneracy are our two main points here}
Assumption (\ref{eq:assumption}) greatly restricts the class of models that can be fit under the GERGM framework. To appropriately fit structural features of a network such as the degree distribution, reciprocity, clustering or assortative mixing, it may be necessary to use network statistics that involve nonlinear functions of the edges. Under Assumption (\ref{eq:assumption}), nonlinear functions of edges are not permitted -- a limitation that may prevent theoretically or empirically appropriate models of networks in many domains.  Furthermore, as we will demonstrate in our numerical study, exponentially weighted network statistics like those in Table \ref{tab:stats} can provide a means to flexibly model networks. This is particularly beneficial in cases where a theoretically appropriate non-degenerate model cannot be identified within the restricted class of GERGMs. To incorporate the aforementioned statistics and extend the class of available GERGMs, we develop a general inferential framework via Metropolis--Hastings that is applicable to any GERGM specification. 

\subsection{A General Inferential Framework via Metropolis--Hastings} \label{sec:MH}

% Consider generating $M$ network samples $x^{(1)}, \ldots, x^{(M)}$. 
An alternative and more flexible way to sample a collection of networks from the density $f_X(x, \bt)$ is the Metropolis--Hastings procedure. The Metropolis--Hastings procedure that we propose samples the $t+1$st network, $x^{(t+1)}$, via a truncated multivariate Gaussian proposal distribution $q(\cdot | x^{(t)})$ whose mean depends on the previous sample $x^{(t)}$ and whose variance is a fixed constant $\sigma^2$. %\footnote{We note that we have also considered the use of a Beta proposal distribution. We discuss this alternate specification in the supplement of this paper.} 

The truncated Gaussian is a convenient and commonly used proposal distribution for bounded random variables such as those on the $[0,1]$ interval with which we are working (see, e.g., \cite{browne2006mcmc,claeskens2010multiresolution,muller2010mcmc,neelon2014spatial,franks2014estimating}). The advantage of the truncated Gaussian over the obvious alternative for bounded random variables -- the Beta distribution -- is that it is straightforward to concentrate the density of the truncated Gaussian around any point within the bounded range. For example, a truncated Gaussian with $\mu=0.75$ and $\sigma=0.05$ will result in proposals that are nearly symmetric around 0.75 and stay within 0.6 and 0.9. In practice, we found the shape of the Beta distribution to be less amenable to precise concentration around points within the unit interval, which leads to problematic acceptance rates in the Metropolis--Hastings algorithm.  
%\todo[inline]{What does adding excess skew or kurtosis mean? Give some other justification (easy to sample, does well in practice etc) or clarify. } 

We say that $w$ is a sample from a truncated normal distribution on $[a,b]$ with mean $\mu$ and variance $\sigma^2$ (written $W \sim \text{TN}_{(a,b)}(\mu, \sigma^2)$) if the pdf of $W$ is given by:

$$g_W(w | \mu, \sigma^2, a,b) = \dfrac{\sigma^{-1}\phi(\frac{w - \mu}{\sigma})}{\Phi(\frac{b - \mu}{\sigma}) - \Phi(\frac{a - \mu}{\sigma})}, \text{\hskip 1pc $a \leq w \leq b$}$$

\noindent where $\phi(\cdot | \mu, \sigma^2)$ is the pdf of a $N(\mu, \sigma^2)$ random variable and $\Phi(\cdot)$ is the cdf of the standard normal random variable. To ease notation, we write the truncated normal density on the unit interval as
% the proposal density $q_{\sigma}(y | x)$:

\begin{equation}\label{eq:proposal}
	q_\sigma(w | x) = g_W(w | x, \sigma^2, 0, 1)
	\end{equation}
This will be our proposal density.
Denote the weight between node $i$ and $j$ for sample $t$ by $x^{(t)}_{ij}$. The Metropolis--Hastings procedure we employ generates sample $x^{(t+1)}$ sequentially according to an acceptance/rejection algorithm. The $t+1$st sample $x^{(t+1)}$ is generated as follows.   

% sample $x^{(t+1)}$ is generated in the following manner.

\begin{enumerate}
	\item For $i,j \in [n]$, generate proposal edge $\tilde{x}_{ij}^{(t)} \sim q_{\sigma}(w | x^{(t)}_{ij})$ independently across edges.
	\item Set\\
	$$x^{(t+1)} = \begin{cases} \tilde{x}^{(t)} = (\tilde{x}^{(t)}_{ij})_{i,j \in [n]} & \text{w.p. \hskip 1pc} \rho(x^{(t)}, \tilde{x}^{(t)})\\
							x^{(t)} & \text{w.p. \hskip 1pc} 1 - \rho(x^{(t)}, \tilde{x}^{(t)}) \end{cases}$$
	\end{enumerate}
	\indent where 
	\begin{align}\rho(x,y) &= \text{min}\left(\dfrac{f_X(y | \bt)}{f_X(x| \bt)} \prod_{i,j\in [n]}\dfrac{ q_{\sigma}(x_{ij}|y_{ij})}{q_{\sigma}(y_{ij} | x_{ij})}, 1\right) \nonumber \\
		&= \text{min}\left(\exp\left(\bt'(\bh(y) - \bh(x))\right) \prod_{i,j\in [n]}\dfrac{ q_{\sigma}(x_{ij}|y_{ij})}{q_{\sigma}(y_{ij} | x_{ij})}, 1\right) \end{align} 
		
The acceptance probability $\rho(x^{(t)},\tilde{x}^{(t)})$ can be thought of as a likelihood ratio of the proposed network given the current network $x^{(t)}$ and the current network given the proposal $\tilde{x}^{(t)}$. Large values of $\rho(x^{(t)},\tilde{x}^{(t)})$ suggest a higher likelihood of the proposal network. It is readily verified that the resulting samples $\{x^{(t)}, t = 1,\ldots, M\}$ form a Markov Chain whose stationary distribution is the target pdf $f_X( \cdot | \bt)$.

The proposal variance parameter $\sigma^2$ influences the average acceptance rate of the Metropolis--Hastings procedure described above. Indeed, the value of $\sigma^2$ tends to be inversely related to the average acceptance rate of the algorithm. \cite{roberts1997weak} analyzed the efficiency of general random walk Metropolis algorithms and found that an acceptance rate of 0.234 optimized the convergence rate of this class of algorithms. Following their heuristic, we suggest tuning $\sigma^2$ so that the average acceptance rate is approximately 0.25.\footnote{We introduce this criterion as a heuristic for MH sampling for GERGM, since the conditions outlined by  \cite{roberts1997weak} for 0.234 to be optimal do not apply to sampling from GERGM.} 

The Metropolis--Hastings algorithm requires specification of an initial sample $x^{(1)}$. To this end, we sample $x^{(1)}$ from a collection of independent uniform random variables on the unit interval. We set a sufficient burn-in so that the resulting chain of $M$ samples have converged. To test the convergence of the samples, we use the Geweke dignostic test for stationarity \citep{geweke1991evaluating} on the network statistics associated with the collection of samples. Furthermore, traceplots of the network statistics can be used to readily surveil the convergence of the network samples. We illustrate how to diagnose convergence in the numerical study in the Appendix.

%Section: Flexible Specification of Network Statistics
\section{Flexible Model Specification} \label{sec:networkstats}
% \todo[inline]{Bruce -- Please re-write this section. Depending on how we choose to specify our statistics with $\alpha$, we need to support our choice (either due to empirical results, or the relationship with GWESP and local dependence). It would be good too to discuss the effects of $\alpha$, for instance what it means when it is 0, or 1. Perhaps this can be shortened, or the table avoided.}
In the context of the dichotomous ERGM, a substantial literature has arisen around how to best formulate network statistics that represent important generative relational processes such as transitivity, balance, and preferential attachment \citep{Wasserman:1996,Park:2004,Snijders:2006,Hunter:2007}. The initial development of ERGM specifications focused on local subgraph counts, such as the number of two-stars and triangles, that implied straightforward conditional distributions for each tie given the rest of the network (i.e., Markov graphs \citep{frank1986markov}). Intermediate extensions of the standard suite of network statistics used in ERGM specifications focused on more advanced or higher-order subgraph counts \citep{Pattison:2002}, reflecting longer paths and clique-like structures among node sets. 

Unfortunately, in most cases, these motif-count specifications lead to empirically implausible models due to the problem of degeneracy. \cite{Snijders:2006} and \cite{Hunter:2007} propose the use of geometrically decreasing weights in the calculation of statistics for transitivity, and for in- and out-degree distributions. The down-weighting in these statistics takes effect as a single node or edge is involved in many subgraph motifs (e.g., the contribution to the transitivity statistic from the first shared partner to two nodes incident to an edge, is more than four times the contribution of the fourth shared partner). These geometrically weighted specifications were shown to avoid degeneracy with much greater success than models specified with simple local subgraph counts. The geometrically weighted shared partners (GWESP) statistics from \cite{Snijders:2006} and \cite{Hunter:2007} reduces the weight of high order statistics in an ERGM and reduces the computational complexity of typical subgraph counting. \cite{Wyatt:2010} suggest using the geometric mean of subgraphs as the measure of ``subgraph intensity'' for network statistics. 

In the GERGM framework, we specify statistics that correspond to the subgraph configurations that have proven fruitful in specifying binary-valued ERGMs. Though virtually any network statistic can be used in a GERGM specification, we focus on a flexible, two-pronged, weighting scheme that dampens the extremes that arise through summed subgraph products. The geometric mean suggested in \cite{Wyatt:2010} can be seen as dampening the change in subgraph sums with respect to subgraph product values by exponentiating the subgraph product to an exponent between 0 and 1. The first prong in our weighted specifications can be considered a generalization of the geometric mean. That is, we suggest exponentiating each sub-graph by exponent $(\alpha \in (0,1])$ before summing over all subgraphs. We refer to this as $\alpha$-inside weighting. The second prong in our specifications represents an extension of the triangle model specification in \cite{lubetzky2015replica}. \cite{lubetzky2015replica} show that raising the triangle density to an exponent greater than zero, but less than 2/3 leads to an ERGM specification that is asymptotically distinguishable from Erdos-Renyi random graphs, which is not true of the conventionally-specified (i.e., non-exponentiated) ERGM statistics. We refer to the latter prong as the $\alpha$-outside specification. 

Aside from providing different empirical fit, the $\alpha$-outside model leads to a more complicated pattern of dependence among the ties, with all ties dependent upon each other, to a degree. The $\alpha$-inside weighting leads to the local dependence common to ERGMs, in which the change statistics (i.e., derivatives of $h$ with respect to edge values in the GERGM) depend upon edges in which an edge is embedded in subgraphs relevant to the statistics. Formally, as long as the statistics being raised to $\alpha$ are sub-graph products, the $\alpha$-inside weighting leads to a Markov graph \citep{frank1986markov} form of the GERGM, in which the joint density of the constrained (i.e., quantile) graph factorizes to a product over functions of sub-graphs. \cite{frank1986markov}, drawing on the Hammersley-Clifford theorem, discuss how ERGM specifications that factorize by sub-graphs exhibit local dependence in which edges depend only on neighbors within the subgraphs. Since it does not factorize by sub-graphs, the $\alpha$-outside specification leads to global dependence, in which the change statistics depend upon the local edges as well as the global network statistic values. This is readily observed by considering the derivative of a statistic weighted according to the $\alpha$-outside specification with respect to a change in an edge $X_{ij}$. Let $h_\alpha(X) = h(X)^{\alpha}$, then   
\begin{equation} \frac{dh_\alpha}{dX_{ij}} = \frac{\alpha}{h^{1-\alpha}}\frac{dh}{dX_{ij}}. \label{eqn:derivative} \end{equation}

We see here that the change statistic with respect to an edge increases with the values of the edges that are local to the edge in a given network statistic (i.e., $\frac{dh}{dX_{ij}}$), but decreases with the global value of the network statistic (i.e., $h^{1-\alpha}$). The decrease with the global value of the statistic is a dampening effect according to which the tendency to form dense motifs lessens with the average/total density of those motifs across the network. We consider these two approaches to dampening the combinatorial growth in network statistic values, and show that each method can be used to avoid degeneracy in GERGM. We note that in principle one can specify any suite of network statistics for a GERGM specification. In this work, we specifically consider $\alpha$-outside specification using the statistics described in Table \ref{tab:stats}. In Section \ref{sec:numerical}, we show that our chosen flexible network statistics provide a means to avoid degeneracy in the GERGM and capture relevant network motifs in application.

\begin{table}[ht]
\centering
\tabcolsep = 0.11cm
{\small
\singlespacing
\begin{tabular}{c c c}
{\bf Network Statistic} & {\bf Parameter} & {\bf Value} \\ \hline
Reciprocity & $\theta_{\text{R}}$ & $\left(\displaystyle\sum_{i < j} x_{ij}x_{ji}\right)^{\alpha_\text{R}}$ \\ \hline
Cyclic Triads & $\theta_{\text{CT}}$ & $\left(\displaystyle\sum_{i < j <k} \left(x_{ij}x_{jk}x_{ki} + x_{ik}x_{kj}x_{ji}\right)\right)^{\alpha_{\text{CT}}}$ \\ \hline
In-Two-Stars & $\theta_{\text{ITS}}$ & $\left(\displaystyle\sum_{i} \sum_{j < k \neq i} x_{ji}x_{ki} \right)^{\alpha_{\text{ITS}}}$ \\ \hline
Out-Two-Stars & $\theta_{\text{OTS}}$ &$\left(\displaystyle\sum_{i} \sum_{j < k \neq i} x_{ij}x_{ik}\right)^{\alpha_{\text{OTS}}}$ \\ \hline
Edge Density & $\theta_{\text{E}}$ &$\left(\displaystyle\sum_{i \neq j}x_{ij}\right)^{\alpha_{\text{E}}}$\\ \hline
Transitive Triads & $\theta_{\text{TT}}$ &$\left(\displaystyle\sum_{i < j < k} \left(x_{ij}x_{jk}x_{ik} + x_{ij}x_{kj}x_{ki} + x_{ij}x_{kj}x_{ik} \right) + \right. $\\
&\hskip .75pc &$\left. \displaystyle\sum_{i < j < k} \left(x_{ji}x_{jk}x_{ki} + x_{ji}x_{jk}x_{ik} + x_{ji}x_{kj}x_{ki}\right)\right)^{\alpha_{\text{TT}}}$\\
\end{tabular}
}
\caption{\small Summary of network statistics used in the specification of a GERGM in this work. These are the $\alpha$-outside specification of five commonly-used network statistics.  \label{tab:stats}}
\end{table}

%Section: Numerical Study
\section{Applications}\label{sec:numerical}
We assess the performance and utility of our proposed Metroplis--Hastings procedure for the GERGM using real and simulated networks. First, we analyze an application in which the Metropolis--Hastings sampler can be used to fit non-degenerate model specifications in a situation where the Gibbs sampler is not available. For this, in Section \ref{sec:lending} we analyze an international lending network that contains the aggregate bank lending volume between 17 large industrialized nations in 2005. In Section \ref{sec:migration} we analyze the U.S. state migration network from 2006 to 2007. In this example, we validate our Metropolis--Hastings procedure by numerically comparing its estimates with those obtained from the Gibbs approach in \cite{Desmarais:2012}. In Section \ref{sec:sims} we explore the utility of flexible model specification for a directed variant of the \emph{two-star model} \citep{handcock2003assessing}. In binary networks, the two-star model is known to be prone to dengeneracy given small changes in its parameter values \citep{park2004solution}. Our simulation study suggests that one can easily identify non-degenerate GERGM specifications for a weighted version of the two-star model. Importantly, we show that under certain weightings, Metropolis--Hastings can simulate networks with any desired edge density and clustering structure. The R code and all of the data used in this section are available in the online supplement.

\subsection{International Lending Network} \label{sec:lending}
% \todo[inline]{Matt -- please re-write. Include: (i) brief description of model, (ii) description of exogeneous statistics and how they relate to international lending, (iii) how $\alpha$ was chosen to find a good fit for MH, (iv) a brief discussion about the hysteresis plots, and (v) a nice picture of the network with a white background that illustrates the weight of the edges through darkness of the edge. Also, the text in the captions of Figures 1, 2, and 3 may need to be altered, but the pictures are up to date.}

Our first application of the GERGM is to the network of aggregate private and public lending between 17 large industrialized nations in 2005. Weighted directed edges between nations represent the total monetary volume, in millions of U.S. dollars, that was loaned from one nation to another. Figure \ref{fig:financial_network_plot} illustrates this weighted network. This data was collected by the Bank for International Settlements (BIS) and a descriptive analysis was originally published in \citet{Oatley2013}. To the best of our knowledge, there have been no published studies of international lending using statistical network models. There are numerous theoretical, exploratory, and descriptive analyses on international lending as a network phenomenon \citep{Niemira2004, Nier2007, Rodriguez2007, Gai2010,Amini2011, Billio2012}, especially in the wake of the 2008 financial crisis. One particular challenge in this network is the heavy tailed nature of the lending volumes (with the majority of lending concentrated between Germany, Great Brittan, Japan, and the United States). We first apply an ln(x+1) transformation on all aggregate lending flows between countries - a standard practice in international finance applications - and subsequently model the transformed edge weights using the GERGM.

%This transformation reduces the skew in the edge weight distribution, but does not eliminate all of the challenges associated with modeling a network with a heavy-tailed degree distribution. 

\begin{figure}[ht]
	\centering
	\includegraphics[width = .6\textwidth, trim = 0cm 0cm 0cm 0cm, clip = TRUE]{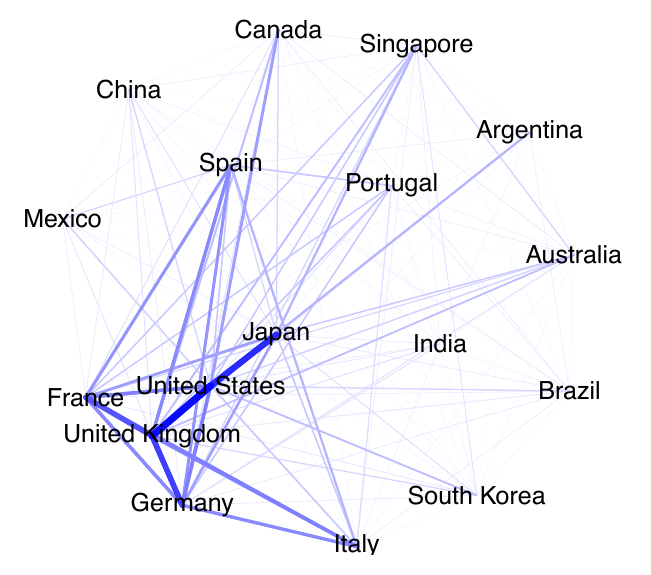}
\caption{\small Network plot of the international aggregate interbank lending network. Darker edges indicate a larger volume of lending. \label{fig:financial_network_plot}}
\end{figure}

We control for several important exogenous predictors in the GERGM specification. In particular, we include sender and receiver effects for the (natural log) gross domestic product (GDP) as we expect countries with larger economies to both lend and borrow more than those with smaller economies. We also include network predictors that represent (normalized) aggregate trade volume between countries, as well as the (normalized) number of inter-governmental organization (IGO) co-memberships. We expect that countries that trade more with one another will also lend more with one another, and that those countries that share a larger number of co-memberships in IGOs will also be more likely to lend more frequently with one another due to their increased diplomatic cooperation. Finally, we include mixing matrices parameterizing the propensity for countries to lend to each other based on G8 membership\footnote{The G8 member countries are Canada, France, Germany, Italy, Japan, Russia, the United Kingdom and the United States.}. We chose $T$ as the cdf of a Student's $t$ distribution with one degree of freedom, whose median is a linear regression on the specified exogenous predictors.

% We leave out the base case of non-G8 to non-G8 lending, and expect that G8 to G8 lending will be more likely, as will G8 to non-G8 lending and non-G8 to G8 lending. 

%Together, we find that these predictors account for a large proportion of the variation in international lending that can be explained by the attributes of countries, meaning that our structural parameter estimates should not simply be picking up artifacts of country characteristics. 

In addition to the exogenous predictors discussed above, we also include structural network predictors, including mutual dyads, transitive triads, and out two-stars statistics in our model. These statistics allow us to test for the presence of mutuality, clustering, and economies of scale in lending, all of which are theoretically important in the international trade \citep{Oatley2013}. Although this specification includes a number of exogenous covariates for control, we find that GERGM model with no $\alpha$ down weighting exhibited degeneracy. To address this degeneracy, we considered an exponentially weighted model using statistics from Table \ref{tab:stats}. We used the Metropolis--Hastings procedure to estimate the GERGM where network predictors were down-weighted by $\alpha_\text{R} = \alpha_{\text{TT}} = \alpha_{\text{OTS}} = 0.8$. The 0.8 value was selected because it was the lagest value for which we could consistently estimate a non-degenerate model across multiple runs of estimation. We optimized the Metropolis-Hastings proposal variance at each step in the estimation process (with a target acceptance rate of 0.25 $\pm$ 0.05) initialized a burn-in of 400,000 full network samples, and then sampled 800,000 networks from which we thinned the resulting sample by keeping every two hundredth network. The average acceptance rate was approximately 0.22. The resulting parameter estimates are given in Table \ref{tab:finance}. To assess convergence of the estimated models, we simulated 800,000 networks and compared the distribution of the mutual dyads, transitive triads, and out two-stars statistics to the observed values in the lending network. Further, we investigated the goodness of fit of our model by comparing the distributions of the simulated and observed in two-stars, cyclic triads, and network density distributions. These results are shown in Figure \ref{fig:financial_gof}.

\begin{table}[ht]
	\centering
	\begin{tabular}{l c}
	 & Parameter\\
	 Statistic & Estimate (s.e.)\\ \hline
	 Transitive Triads & 1.387 (0.478) \\
	 Out Two Stars & -2.645 (0.811)\\ 
	 Mutual Dyads & 7.023 (2.900) \\
	 G8 Sender, G8 Receiver & 1.505 (0.186) \\
	 G8 Sender, Non-G8 Receiver & 0.804 (0.177) \\
	 Non-G8 Sender, G8 Receiver & 1.480 (0.147)\\
	 IGO Co-members & 1.390 (0.210)\\
	 log(GDP) Sender & 0.606 (0.126)\\
	 log(GDP) Receiver & 4.887 (1.040)\\
	 Normalized Net Exports & -0.068 (0.046)\\ 
		
	\end{tabular}
	\caption{\small Estimates of the network parameters of the GERGM model when fit to the international lending network via the Metropolis--Hastings procedure. \label{tab:finance}}
\end{table}

\begin{figure}[ht]
\centering
\includegraphics[width = 0.65\textwidth, trim = 0cm 0cm 0cm 3cm]{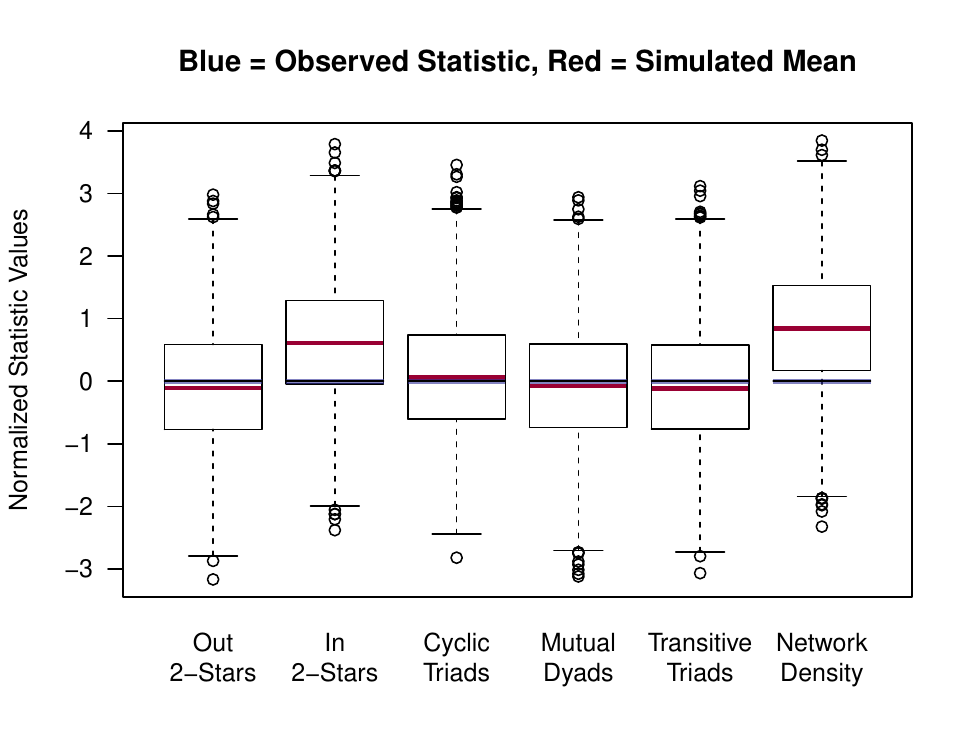}
\caption{\small Convergence and goodness of fit plots for the model fitted to the international lending network. For each fit model, 800,000 networks were simulated using the Metropolis--Hastings sampling procedure and the final exogenous and structural parameter estimates. Each box plot compares the quantiles of simulated networks (with mean statistic values at the red line) to the observed network statistic (blue line). Here, the In 2-Stars and Network Density statistics were not included in the fitted model. \label{fig:financial_gof}}
\end{figure}

%Hysteresis plot
\begin{figure}[ht]
	\centering
	\includegraphics[width = \textwidth, trim = 0cm 0cm 0cm 0cm, clip = TRUE]{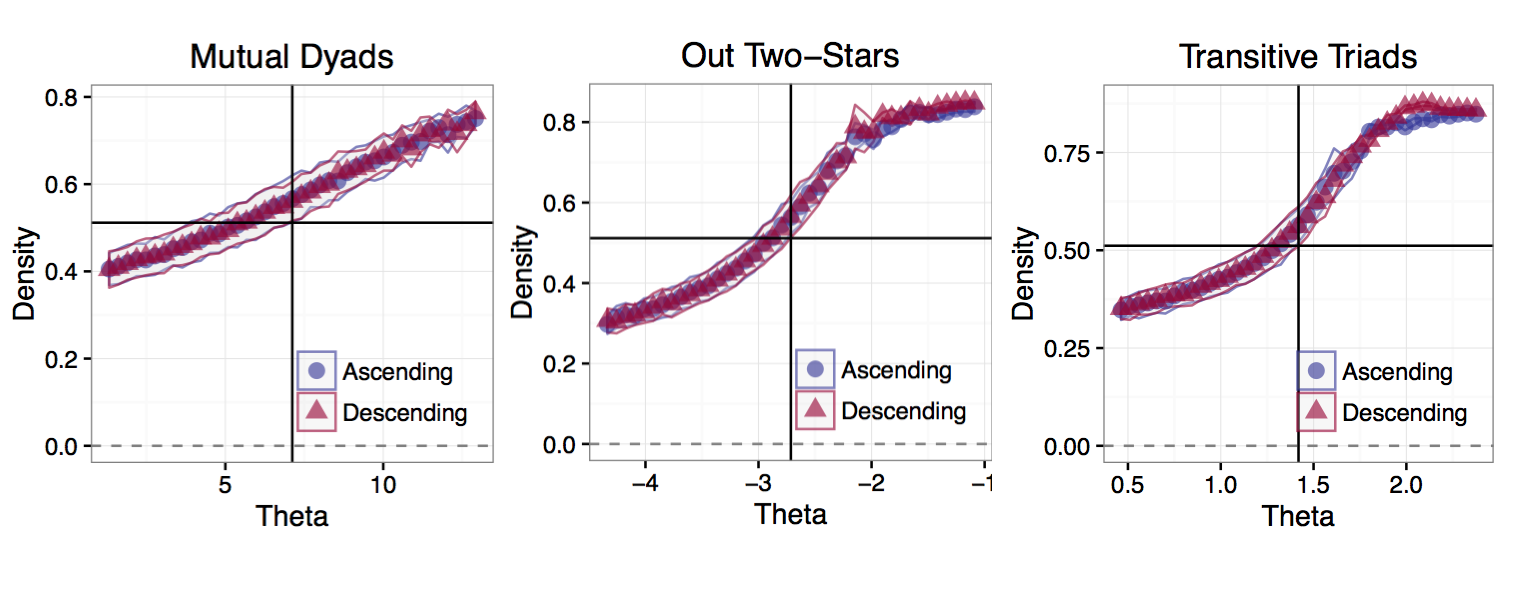}
	\caption{\small Hysteresis plots for the international trade network. Shaded regions cover $\pm 1.96$ standard deviations in the simulated network densities for that parameter value. The vertical black line indicates the parameter value from the main estimation step and the horizontal black line indicates the observed network density.  \label{fig:hysteresis_trade}}
	
\end{figure}

As we can see from Figure \ref{fig:financial_gof}, our model has appeared to have converged based on the distributions of the transitive triads, mutual dyads, and out two-stars statistics. In terms of goodness of fit, we see that our model provides a very good fit for the observed network in terms of cyclic triads; however, the in-two-stars and network density values are somewhat overestimated. Geweke statistics and trace plots of the network density for 800,000 networks simulated from the fitted GERGM specification via Metropolis--Hastings simulation also indicate that the model has converged (see Figure \ref{fig:financial_trace} in the Appendix). The exogenous covariate parameter estimates from our model largely conform to our theoretical expectation, although it is interesting that we see a small (negative) parameter estimate for the effect of trade, indicating that there is not a particularly strong relationship between trade and lending in our network, when controlling for other economic and network factors. We observe positive and statistically significant transitivity and reciprocity effects, and a negative out-two stars effect. These results are interesting because previous studies, including \citet{Oatley2013}, have argued that the international lending network is hierearchical - a property that does not match our results as we would expect to see a positive out two-stars parameter estimate. Further exploration of this finding is outside of the scope of this discussion, but should be considered in future research.

To further asses the potential for degeneracy in our model, we performed a hysteresis analysis similar to that described in \citet{Snijders:2006} for each structural parameter estimate. Starting with a sparse network and holding all other parameter estimates at their posterior means, we varied each structural parameter estimate ten standard deviations below to ten standard deviations above its posterior mean and simulated 500,000 networks using our Metropolis--Hastings procedure from each parameter value combination (with a burnin of 500,000). We changed the structural parameter value 0.5 standard deviations at each iteration of this process, for a total of 41 parameter values. For each new value of the parameter, we used the final network from the M-H simulation using the previous parameter value as the initialization for the M-H simulation for the new specified parameter. We plotted the mean network density against the parameter values in order to asses the potential for jumps in the network density that might indicate an underlying issue with model degeneracy. Figure \ref{fig:hysteresis_trade} shows the hysteresis plots for our model, and these plots do not indicate any obvious issues with degeneracy in this specification.

%Subsection: Application to US Migration Network
\subsection{U.S. Migration Network}\label{sec:migration}
We next apply the GERGM to the U.S. migration network analyzed in \cite{Desmarais:2012}. We note that this application is used for validation of our Metropolis--Hastings procedure; indeed, we compare the estimates obtained with Metropolis--Hastings directly with the estimates obtained from the Gibbs sampler for the same GERGM specification.

Historically, interstate migration has played an important role in the understanding of local financial markets, public infrastructure, and the political climate within each state \citep{clark1981demand, levine1999empirical, gimpel2001interstate}. The network that we model contains 51 nodes that represent the 50 U.S. states as well as Washington, D.C. Directed edges are placed between states in which there was a change in interstate migration flow from 2006 and 2007. The weight, $y_{ij}$, associated with the directed edge from node $i$ to node $j$ is the total change in interstate migration from state $i$ to state $j$ between 2006 to 2007. This data set also contains ten demographic exogenous predictors that further describes the pairwise relationships between states. The predictors describe the geographic distance, and the sender and receiver effects of high January temperature, income, unemployment, and population of the states. Like the application in Section \ref{sec:lending}, we chose $T$ as the cdf of a Student t distribution with one degree of freedom, whose median is a linear regression on the specified demographic predictors.
% %% Choice of model
% We apply the GERGM to this data by specifying $t_{ij}$ as a Cauchy pdf for each edge $(i,j)$ whose median, $\mu_{ij}$, is a linear function of the exogenous predictors:
% \vskip - 1pc
% %Transformation function fo the model
% \begin{align*}
% t_{ij}(y,z, \bm{\beta}) &= {\left(\pi\left(1 + (y_{ij} - \mu_{ij})^2\right)\right)}^{-1},\\
% \mu_{ij} &= \beta_0 + \sum_{k = 1}^{10} \beta_k z_{ij}(k)
% 	\end{align*}

%The vector $\bm{\beta} \in \mathbb{R}^{11}$ parameterizes the transformation of each pair of states through a linear regression on ${\bf z}_{ij}$. 
We incorporated network statistics that represent reciprocity, cyclic triads, in-two-stars, out-two-stars, and transitive triads in our GERGM specification, and following the model fit in \cite{Desmarais:2012} we used no $\alpha$ down-weighting. 

% Our model specification is described by (\ref{eq:fofx}) and (\ref{eq:fofy}) where $\bt = (\theta_{\text{R}}, \theta_{\text{TT}}, \theta_{\text{CT}}, \theta_{\text{OTS}}, \theta_{\text{ITS}})'$, and $\bm{\alpha} = (1,1,1,1,1)$.

%We note that the model used here is the same non-degenerate specification described in \cite{Desmarais:2012} and that Gibbs sampling is applicable. 

We fit the above model using both the Metropolis--Hastings sampling procedure and Gibbs. For Gibbs, we use 50,000 simulated networks with a set burn-in of 10,000 networks in each iteration. We optimized the Metropolis-Hastings proposal variance at each step in the estimation process (with a target acceptance rate of 0.25) initialized a burn-in of 1,000,000 full network samples, and then sampled 2,000,000 networks from which we thinned the resulting sample by keeping every one thousandth network. The average acceptance rate was approximately 0.24. The parameter estimates and associated standard errors for each method are shown on in Table \ref{tab:migration}.

\begin{table}[ht]
	\centering
	\begin{tabular}{l c c}
	 & M-H Parameter & Gibbs Parameter\\
	 Statistic & Estimate (s.e.) & Estimate (s.e.)\\ \hline
	 Transitive Triads & 0.074 (0.053)& 0.078 (0.053)\\
	 Cyclic Triads & -0.206 (0.042)& -0.204 (0.040)\\
	 Out Two Stars & 0.017 (0.044)& 0.011 (0.043)\\ 
	 In Two Stars & -0.029 (0.039)& -0.030 (0.040)\\
	 Mutual Dyads & -0.131 (0.348) & -0.107 (0.338)\\
	 Unemployment Sender & 27.163 (13.481) & 27.402 (13.463)\\
	 Unemployment Receiver & -3.673 (12.382) & -3.475 (12.476) \\
	 Mean January Temp. Sender & -11.031 (14.474) & -11.167 (14.452)\\
	 Mean January Temp. Receiver & -15.147 (13.609) & -15.101 (13.713)\\
	 Population Size Sender & 1.806 (20.264)& 1.744 (20.244)\\
	 Population Size Receiver & -35.532 (16.127) &-35.282 (16.215) \\
	 Mean Income Sender & 2.349 (11.613)& 2.220 (11.583)\\
	 Mean Income Receiver & -1.129 (10.652) & -0.969 (10.735)\\
	 Distance & 7.081 (11.917)& 7.218 (11.970)\\ 
	 Dispersion & 5.942 (0.029) & 5.942 (0.029)\\
		
	\end{tabular}
	\caption{\small Estimates of the network parameters of the GERGM model when fit to the U.S. migration network via the Metropolis--Hastings and Gibbs sampler procedures. \label{tab:migration}}
\end{table}

% \begin{figure}[ht]
% 	\centering
% 	\includegraphics[width = \textwidth]{Images/Migration_Coeffs.png}
% 	\caption{\small Estimates of the network parameters of the GERGM model when fit to the U.S. migration network. Shown are the results for the Gibbs and Metropolis--Hastings procedures. Lines represent 90 and 95$\%$ confidence intervals for each estimate. \label{fig:Migration_ests}}
% \end{figure}

Table \ref{tab:migration} reveals that the Metropolis--Hastings and Gibbs procedures provide comparable estimates for each of the modeled predictors. This suggests that each method simulates from the same distribution, as expected. Furthermore, these fitted GERGM reveals three interesting, perhaps expected, trends in the data: (i) there was increased migration away from states with high unemployment, (ii) there was decreased migration to states with a large population, and (iii) there was decreased migration to states with high unemployment. See \cite{Desmarais:2012} for a more detailed discussion of these results. We provide further estimation diagnostics for the Metropolis--Hastings procedure in the Appendix. 

%All three estimation methods identified eight statistically significant covariates at the 95$\%$ level. Four of these significant covariates were topological: transitive triads, out-two-stars, in-two-stars, and cyclic triads. 

% To assess the convergence of the Metropolis--Hastings simulation procedure, we use the Geweke dignostic test for stationarity \citep{geweke1991evaluating}. We test the convergence of the edge weight, reciprocity, and transitive triads statistics for the 10000 simulated networks from the M-H procedure. For each statistic, we fail to reject convergence according to the Geweke two-sided test that tests the equivalence of the first and last third of the samples (p-values = 0.86, 0.91, 0.84, respectively).

%%Subsection: Simulation Study 
\subsection{Non-Dengenerate Specifications of the Two-Star Model}\label{sec:sims}
% \todo[inline]{JW: focus here should be on the fact that the GERGM does not appear to be degenerate for any value of $\alpha$. She the corresponding histograms and plots! Allude to further research here for better understanding of what's going on here.}
In our simulation study, we consider fitting a GERGM to a directed and weighted variant of the two-star model. Consider an edge configuration $x \in [0,1]^m$. We model the occurrence of $x$ as a function of its edges and in-two-stars:

\begin{equation}\label{eq:twostars} f_X(x, \bt, \alpha) = \dfrac{\exp\left(\theta_E h_E(x) + \theta_{\text{ITS}} h_{ITS}(x, \alpha)\right) }{C(\theta_E,\theta_{\text{ITS}})}, \hskip 1pc \text{$x \in [0,1]^m$} \end{equation} 
	\begin{align*}
		h_E(x) &= \sum_{i \neq j} x_{ij} / m\\
		h_{ITS}(x, \alpha) &= \left(\sum_{i} \sum_{j < k \neq i} x_{ji}x_{ki}\right)^\alpha,
	\end{align*}
\noindent where $C(\theta_E, \theta_{\text{ITS}})$ is the normalizing constant that ensures $f_X(x,\bt)$ integrates to one, $h_E(x)$ is the edge density of $x$, and $h_{ITS}(x,1)$ is the in-two-star density of $x$. When $\alpha = 1$, model (\ref{eq:twostars}) is the directed and weighted version of the two-star model considered in \cite{handcock2003assessing}. Model (\ref{eq:twostars}) is closely related to the triangle model from \cite{jonasson1999random, haggstrom1999phase} and the widely used Ising model for lattice processes. We will refer to model (\ref{eq:twostars}) as the weighted in-two-stars model.

The unweighted two-star model is a well-known example that suffers from likelihood degeneracy (see \citet{handcock2003assessing} or \citet{Snijders:2006} for instance). In this simulation study, we empirically analyze model (\ref{eq:twostars}) following a similar study as that described in \citet{Snijders:2006}. We find that, surprisingly, the weighted in-two-stars model does \emph{not} demonstrate the typical signs of degeneracy. We now describe the simulation study and our findings. 

We first fix the edge density parameter $\theta_E$ at -2 and a value of $\alpha$ between 0 and 1. We then simulate one million size 10 networks following model (\ref{eq:twostars}) for each integer value of $\theta_{\text{ITS}}$ between -10 and 10 using the MH sampler. We calculate the mean edge density and the mean in-two-stars value from the million samples at each value of $\theta_{\text{ITS}}$. We repeat this procedure for $\alpha$ values of 0.10, 0.25, 0.50, 0.75, 0.90, and 1. The results are reported in Figure \ref{fig:simplots}. 

%Figure: transition plots for simulation
\begin{figure}[ht]
	\centering
	\includegraphics[width = \textwidth, trim = 0cm 0cm 0cm 3cm, clip = TRUE]{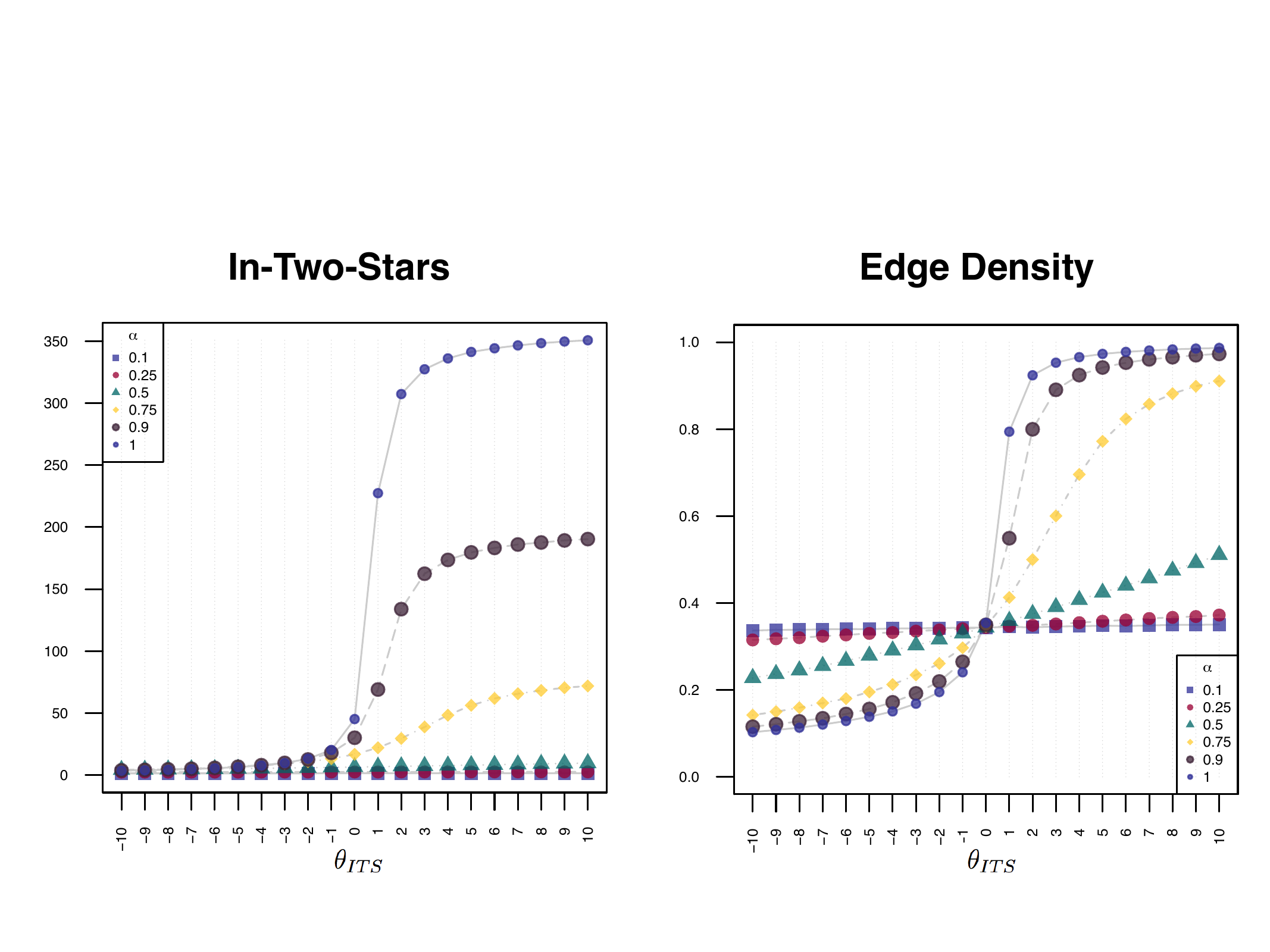}
	\caption{\small The mean value of the in-two-stars statistic and the edge density of one million simulated networks of the in-two-stars model with $\theta_E = -2$ and integer values of $\theta_{\text{ITS}}$ from -10 to 10. The mean values are shown for $\alpha$ values of 0.10, 0.25, 0.50, 0.75, 0.90, and 1.\label{fig:simplots}}
	
\end{figure}

We see from Figure \ref{fig:simplots} that for $\alpha = 1$, there is a large jump in the value of the in-two-stars and edge density statistics between $\theta_{\text{ITS}} = 0$ and $\theta_{\text{ITS}} = 1$. As $\alpha$ decreases, the relative magnitude of this jump decreases and the statistics' curves are relatively smooth over changes in $\theta_{\text{ITS}}$. When $\alpha$ is too small ($\leq 0.50$), the in-two-stars statistic value approaches one and the edge density only changes slightly across values of $\theta_{\text{ITS}}$. The jump phenomenon was also witnessed for binary networks in the two-stars model in \citet{Snijders:2006}, where it was observed that this model specification was most prone to degeneracy issues.  As a consequence, one may expect the empirical distribution of the in-two-stars and edge-density statistics in the neighborhood of $\theta_{\text{ITS}} = 0$ and $\theta_{\text{ITS}} = 1$ to be bimodal at large values of $\alpha$. 

To investigate whether this is the case, we performed a more fine grained grid search for the value of $\theta_{\text{ITS}}$ at which the edge density of the network changed the most, for each value of $\alpha$. We found that, for example, when $\alpha = 0.5$, the steepest change in the edge density occurred at approximately $\theta_{ITS} = 0.55$. Similarly, for $\alpha = 0.75$ and $\alpha = 1$, the steepest changes in the edge density occurred at approximately $\theta_{ITS} = 0.65$ and $\theta_{ITS} = 0.75$, respectively.  We show the empirical distribution for both of the statistics for $\alpha$ values of 0.50, 0.75, and 1.00 at the values of $\theta_{\text{ITS}}$ given above, in Figure \ref{fig:alpha1}. Figure \ref{fig:alpha1} suggests that these distributions are \emph{not} bimodal for these values of $\alpha$, including $\alpha = 1$. We also evaluated these distributions for all other values of $\theta_{\text{ITS}}$ in our simulations and found similar results. Furthermore, these results are not sensitive to the value of $\theta_E$, as it serves only to shift the curves depicted in Figure \ref{fig:simplots} left or right. These findings suggest that the weighted in-two-stars does not suffer from the same degeneracy issues as its binary counterpart.  

%Figure: histograms of unweighted specification statistics
\begin{figure}[ht]
\centering
\includegraphics[width = .9 \textwidth, trim = .1275cm 0cm .5cm 0cm, clip = TRUE]{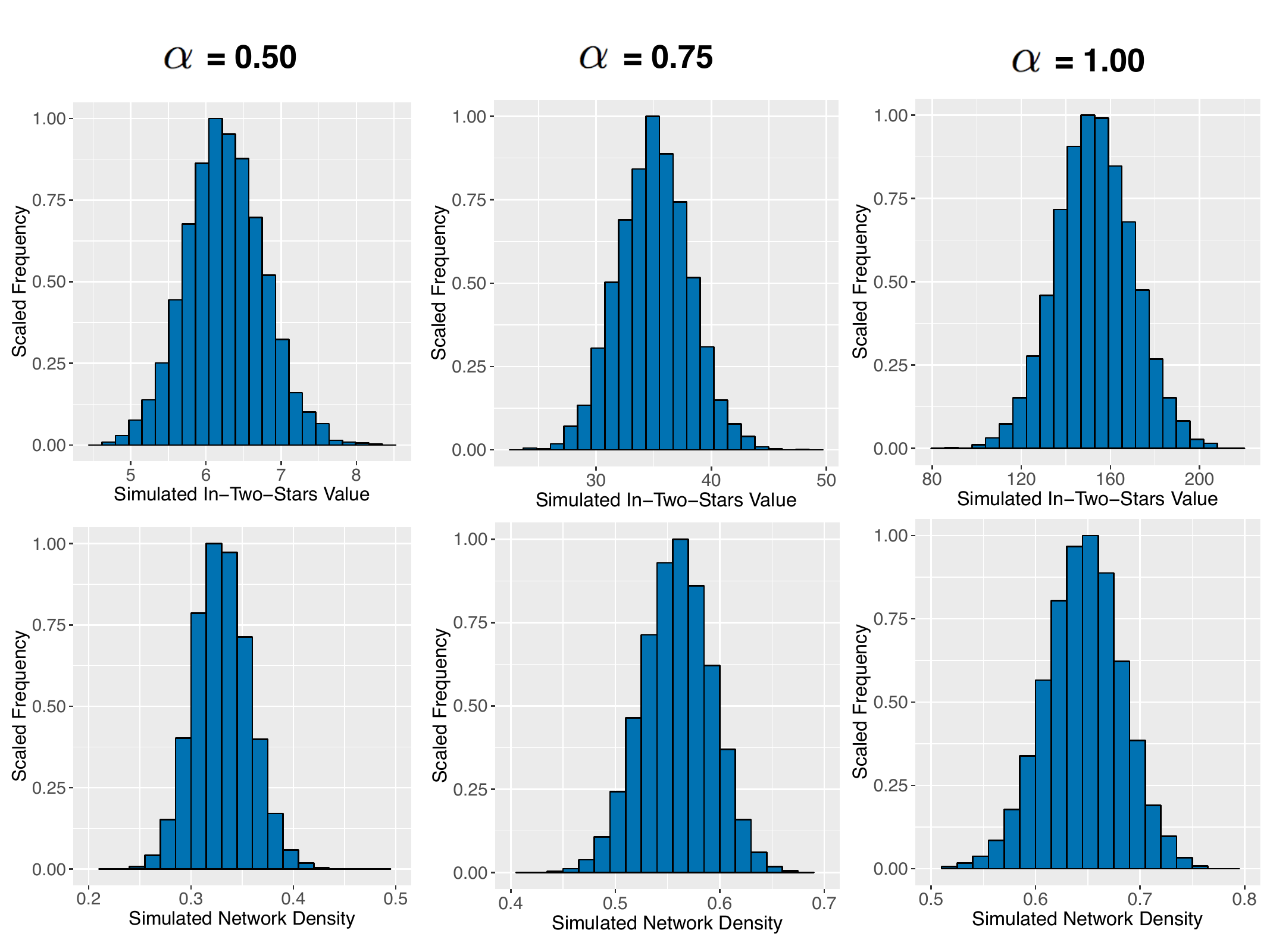}
\caption{\small The empirical (scaled) frequency distribution of the edge density and the in-two-stars value for the in-two-stars model at $\theta_{\text{ITS}} = \{0.55,0.65,0.75\}$ (from left to right). One million networks were simulated and the values for each statistic is shown over all networks. \label{fig:alpha1}}
\end{figure}
%%%%%
In summary, this simulation study provides insights into two important features of the GERGM specification of the two-stars model. First, the weighted in-two-stars models does not appear to suffer from degeneracy at any value of $\alpha$. This surprising result is contrary to the well-known unweighted two-stars model. This simulation also gives some intuition as to how to choose the tuning parameter $\alpha$. Small values of $\alpha$ ($\leq 0.50$) dampened the effect of the in-two-stars statistic too drastically and are therefore not suggested. We encourage using values of $\alpha$ between 0.5 and 0.9 as these values lead to decreased sensitivity of the GERGM model to parameter changes.  

\section{Discussion} \label{sec:discussion}

We have proposed, explicated, and demonstrated several advances in the statistical modeling of weighted networks by substantially increasing the utility of the GERGM. These extensions to the GERGM, taken together, represent a significant increase in the model's capabilities, such that it is now possible to use nearly any model specification for inference on continuous-valued weighted graphs.

% HM contribution 
First, we have proposed and implemented a Metropolis--Hastings algorithm for fitting GERGMs. In the original development of the GERGM, \citet{Desmarais:2012} proposed a Gibbs sampling strategy to estimate the model. However, this approach is limited by the fact that fairly strict constraints are placed on the set of network statistics that can be used in the model. Our Metropolis--Hastings procedure relaxes these restrictions and allows one to use the full set of possible specifications for the model. 

% alpha weighted contribution
Second, we have proposed an approach to dampening the extreme values often produced by subgraph sums and thus avoiding model degeneracy. This dampening technique, because it is critical in avoiding degenerate model specifications in the GERGM, allows analysts to specify a practical and diverse set of endogenous effects as part of the network data generating process. Though this may seem a simple extension to the means by which statistics are computed on the network, this weighting strategy is important because degeneracy is a major obstacle to estimation of inferential models on real-world networks. 

%interpretability of alpha-outside vs. alpha-inside. 
We consider two approaches to network statistic dampening---one in which subgraph-specific sums are raised to a fractional exponent (i.e., $\alpha$-inside dampening), and a stronger approach that involves raising the sum over all subgraphs to a fractional exponent (i.e., $\alpha$-outside dampening). It is important to re-iterate that, while the $\alpha$-inside approach conforms to the local dependence that is typical to ERGM formulations, the $\alpha$-outside approach induces global dependence in that each tie variable depends to some degree on the value of every other tie variable in the network. We see from Equation \ref{eqn:derivative} that the $\alpha$-outside formulation exhibits a form of dependence similar to the $\alpha$-inside formulation in that high edge values are more likely if they contribute to local configurations that are themselves high and associated with positive parameter values. However, the $\alpha$-outside formulation exhibits an additional form of dependence in that the likelihood of high edge values embedded in high value configurations decreases as the global sum over the respective configuration type increases. This global dampening in the $\alpha$-outside model is inversely related to the value of $\alpha$. Though the $\alpha$-inside and $\alpha$-outside formulations may both be considered in efforts to avoid degeneracy, we note that researchers should also consider how the choice between these two formulations affects the interpretation of results.

% Extensions: Model selection and specification
Though we have presented important innovations here, much work remains.  Specifically, we have just scratched the surface when it comes to model selection and specification for GERGMs. First, both in \cite{Desmarais:2012} and in the current study, the statistics used to specify the GERGM have represented straightforward functional adaptations of the statistics commonly used for binary ERGMs. Future research should consider suites of statistics that are applicable to the special case of weighted networks. Furthermore, our approach to weighting the subgraph products requires a choice of $\alpha$ that will rarely be theoretically informed. In our simulation study, we analyzed the effects of $\alpha$ on the sensitivity of the two-stars model and found encouraging results for $\alpha \in (0.5, 1]$. In principle, one could use an alternative data-driven approach that chooses $\alpha$ based on goodness of fit summaries. We plan to investigate this more fully in future work. Finally, the results in the simulation study gave empirical evidence in one well-studied model that the GERGM does not suffer from degeneracy like its binary ERGM counterpart. We aim to theoretically formalize these findings in future work.

\section*{Appendix}
\subsection*{A: Pseudo-code for MCMC Maximum Likelihood Estimation of the GERGM}
\singlespacing
%Inserted Algorithm
\begin{algorithm}[H]
{\footnotesize		
\caption{ {\scshape GERGM MCMCMLE}}
 \SetAlgoLined
 \KwData{Data and parameters}
y $\leftarrow$ vector of edge weights\\
Tol1 $\leftarrow$ tolerance level for GERGM estimation convergence\\
Tol2 $\leftarrow$ tolerance level for Metropolis--Hastings algorithm convergence\\
N $\leftarrow$ number of network samples generated for each iteration of Metropolis--Hastings sampler\\
%M $\leftarrow$ number of network statistics\\
h $\leftarrow$ vector of functions to calculate network statistics\\
T $\leftarrow$ transformation function\\
c $\leftarrow$ shape parameter \com[Default = 1]\\
t $\leftarrow$ 0 \com[iteration number]\\
$\Delta_1$ $\leftarrow$ 1000\\
$\Delta_2 \leftarrow$ 1000\\
%$\Lambda_0 \leftarrow  $ M $\times 1$ vector of 0's\\
%$\theta_0 \leftarrow  $ M $\times 1$ vector of 0's\\ 
 
\bigskip

\While{$\Delta_1 >$ \emph{Tol1}}{
  t++;  \com[Increment iteration]\\
\vskip 1pc
\eIf{t = 1}{{
\scshape Estimate $\beta_{t}$} via MPLE}{
  {\scshape Estimate $\beta_{t}$} via Gradient Descent
  \vskip 1pc
  {\scshape Estimate Theta} via {\bf Algorithm 2} to obtain $\theta_{t}$}
\vskip 1pc
\eIf{$\|\beta_{t-1}\|_{2}^2 \|\theta_{t-1}\|_{2}^2 > 0$}{
$\Delta_1 = \frac{1}{2}\left[\dfrac{\|\beta_{t} - \beta_{t - 1}\|_{2}^2}{\|\beta_{t-1}\|_{2}^2} +  \dfrac{\|\theta_{t} - \theta_{t - 1}\|_{2}^2}{\|\theta_{t-1}\|_{2}^2}\right]$} 
{$\Delta_1 = \frac{1}{2}\left[\|\beta_{t} - \beta_{t - 1}\|_{2}^2 +  \|\theta_{t} - \theta_{t - 1}\|_{2}^2\right]$}
  
} % end while

\KwRet{\emph{($\beta_t,\theta_t$)}}
}
\end{algorithm}

\pagebreak
%%%JW: Starting here 11/18/14
%\pagebreak
\begin{algorithm}[H]
		{\footnotesize
\caption{ {\scshape Estimate Theta}}
 \SetAlgoLined
\com[Initialize parameters]\\

$\hat{x} = T(y,\beta_t)$ \com[current estimate of edge weights based on estimate of $\beta_t$] \\
$\tilde{\theta}_0 = \theta_{t-1}$ \com[initialization of coefficient vector]\\
r = 0 \com[iteration number]\\
m = length(y) \com[Number of edges]\\
\vskip 1pc
\While{$\Delta_2 >$ \emph{Tol2}}{
	r++ \com[Increment iteration]\\
	\vskip 1pc
	%\com[Simulate initial sample $x_0$]\\
	Simulate a sample $x_0$ of length $m$ from density:\\
	$$f(x | \tilde{\theta}_{r-1}) = \dfrac{\exp[\tilde{\theta}_{r-1}' h(x)]}{C(\tilde{\theta}_0)}, \hskip 2pc \text{$C(\tilde{\theta}_{r-1}) = \int_{[0,1]^M} 	\exp[\tilde{\theta}_{r-1}' h(z)] dz$}$$\\
\vskip 1pc
	{\scshape Run Metropolis Hastings Update} via {\bf Algorithm 3} to obtain samples $x_1, \ldots, x_N$.\\
\vskip 1pc

	\com[Update $\tilde{\theta}_r$]\\
	$\tilde{\theta}_{r} = \text{argmax}_{\theta}\left[\theta'h(\hat{x}) - \log[\widehat{C}(\theta)]\right]$ via Gradient Descent, where\\
	%\com[$\widehat{C}(\theta)$ above is a function of $\theta$ given by the following equation:]\\
	$\widehat{C}(\theta) = \dfrac{C(\tilde{\theta}_{r-1})}{N} \displaystyle\sum_{j = 1}^N \exp[(\theta - \tilde{\theta}_{r-1})'h(x_{j})]$\\
	% \com[The new samples $x_1, \ldots, x_N$ were used to estimate this normalizing equation]
% 	\vskip 1pc
	\com[Calculate distance from previous step]\\
	\eIf{$\|\tilde{\theta}_{r-1}\|_{2}^2 > 0$}
	{$\Delta_2 =  \dfrac{\|\tilde{\theta}_{r} - \tilde{\theta}_{r - 1}\|_{2}^2}{\|\tilde{\theta}_{r-1}\|_{2}^2}$}
	{$\Delta_2 =  \|\tilde{\theta}_{r} - \tilde{\theta}_{r - 1}\|_{2}^2$}
}
\KwRet{\emph{$\tilde{\theta}_r$}}
}
\end{algorithm}

\pagebreak

\begin{algorithm}[H]
{\footnotesize
\caption{ {\scshape Metropolis Hastings Update }}
\SetAlgoLined
\KwData{Data and Parameters}
%$N$ : number of network samples to generate\\
%$h$ : Network statistics indicator vector (tells us which ones to use)\\
$\sigma^2$ : variance of Truncated Normal density \com[Default = 1]\\
$x^0 $ Initial weighted network sample (all edges have values between 0 and 1.\\
$n$ : total number of nodes in the network

\bigskip
\com[loop over number of MH network samples]\\
\For{k = 0; k $<$ \emph{N}; k++}{

   		\bigskip
               	\com[Draw a single proposal sample]\\

\For{i = 0; i $<$ \emph{n}; i++}{ 

\For{j = 0; j $<$ \emph{n}; j++}{ 
            % \com[Inside this loop we draw a single new edge weight and calculate both  probabilities. Make sure to ignore self edges which are always fixed at zero]\\
Sample a proposed edge weight $w_{ij}$ from a truncated normal distribution centered at the previous iteration's edge weight
\begin{equation*}
w_{ij} \sim TN \left(x_{ij}^{k-1}, \sigma^2\right)
\end{equation*}

Calculate the probability of the proposed edge weight $w_{ij}$ under the truncated normal distribution centered around the current edge weight ($x_{ij}^{k-1}$). 
\begin{equation*}
p(w)_{ij}  = \mathbb{P} \left(w \leq w_{ij} \mid x^{k-1}_{ij}, \sigma^2 \right)
\end{equation*}

Calculate the probability of the current edge weight ($x_{ij}^{k-1}$) under a truncated normal distribution centered around the proposed edge weight ($w_{ij}$). 
\begin{equation*}
p(x)_{ij}  = \mathbb{P}\left(w \leq x_{ij}^{k-1} \mid w_{ij}, \sigma^2 \right)
\end{equation*}
}
}
Set $w$ as the proposed network and $x_{k-1}$ as the previous iteration's network.\\

Calculate the following log joint pdfs:
\begin{align*}
P_x &= \sum\limits_i^n \sum\limits_j^n  \log(p(x)_{ij}) \\
P_w &= \sum\limits_i^k \sum\limits_j^k \log(p(w)_{ij})
\end{align*}
% \com[We end up with just two scalars that we are going to take the ratio of below. So the first part jut subtracts (cause we are in log space). The second part multiplies theta parameters times the difference between --say number of triangles-- the statistics for the new graph  minus the statistics for the old graph and then adds them up. So a negative theta time some quantity that grew in the new graph is going to be bad for the relative likelihood that new graph]
Calculate $\alpha = (P_x - P_w) + \theta' \left[ h(w_{ij}) - h(x_{ij}^{k-1})\right]$\\
		Sample $u$ from a Uniform(0,1) density\\
		\eIf{$log(u) \leq \alpha$}{
			$x_k = w$}{
			$x_k = x_{k-1}$}	
 	}
\KwRet{$x_1, \ldots, x_N$}
}
\end{algorithm}
%%%Algorithm end

%\newpage

\subsection*{B: GERGM Fit Diagnostics}
To evaluate convergence of the Metropolis--Hastings procedure on the international lending network and the U.S. Migration data, we evaluate the trace plot for the network density of the simulated networks over 800,000 simulated networks. We show this plot in Figure \ref{fig:financial_trace}. For the U.S. Migration data, we simulated 100,000 networks with 10,000 burn-in. The modeled statistics, as well as the MCMC traceplot for the network density for this data are shown in Figure \ref{fig:gof}. Visual inspection as well as the Geweke convergence test statistic suggest that the sampler has converged.

%Traceplot for International lending network
\begin{figure}[ht]
\centering
\includegraphics[width = 0.55\textwidth, trim = 2cm 0cm 0cm 0cm]{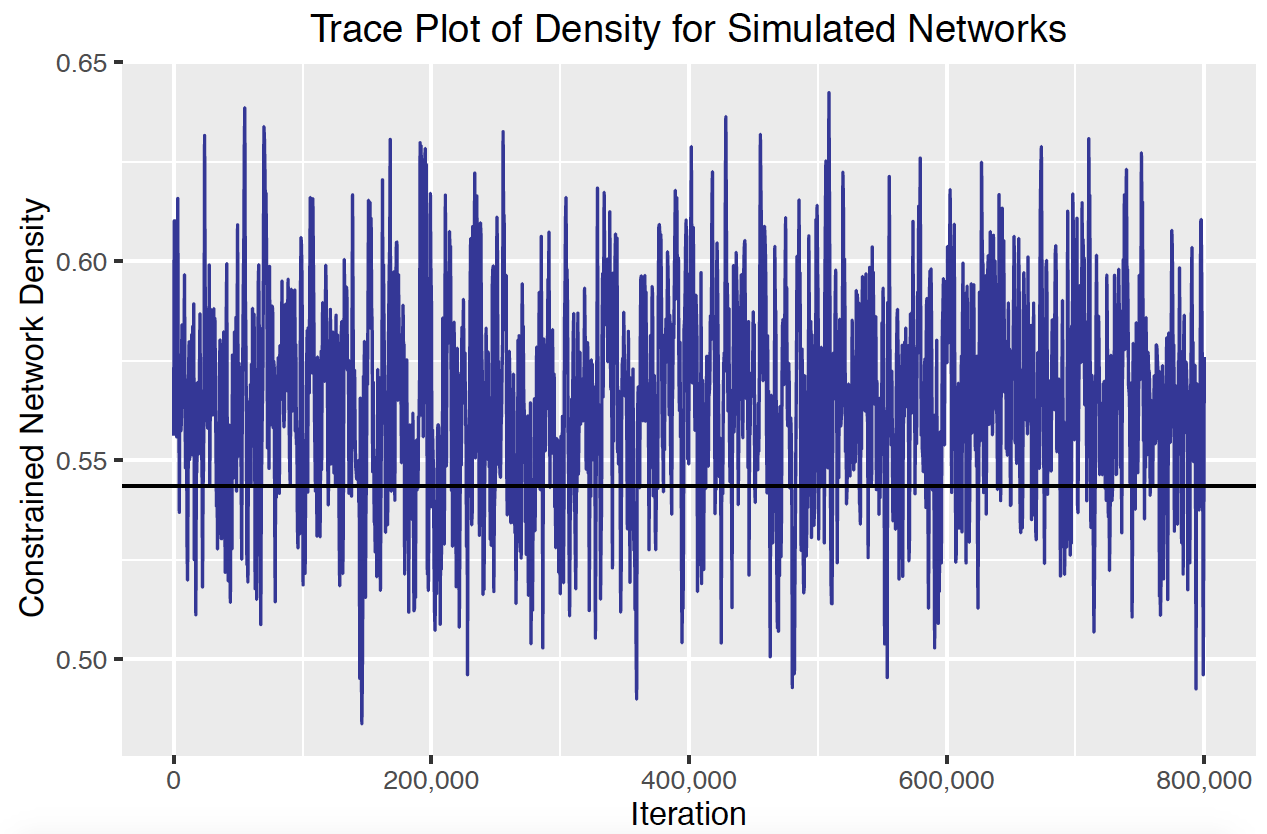}
\caption{\small International lending network trace plot for the network density of the simulated networks over 800,000 simulations from Metropolis--Hastings. \label{fig:financial_trace}}
\end{figure}

%Estimation diagnostics for Migration data
\begin{figure}[ht]
\centering
\includegraphics[width = 0.75\textwidth, trim = 0cm 0cm 0cm 11.5cm, clip = TRUE]{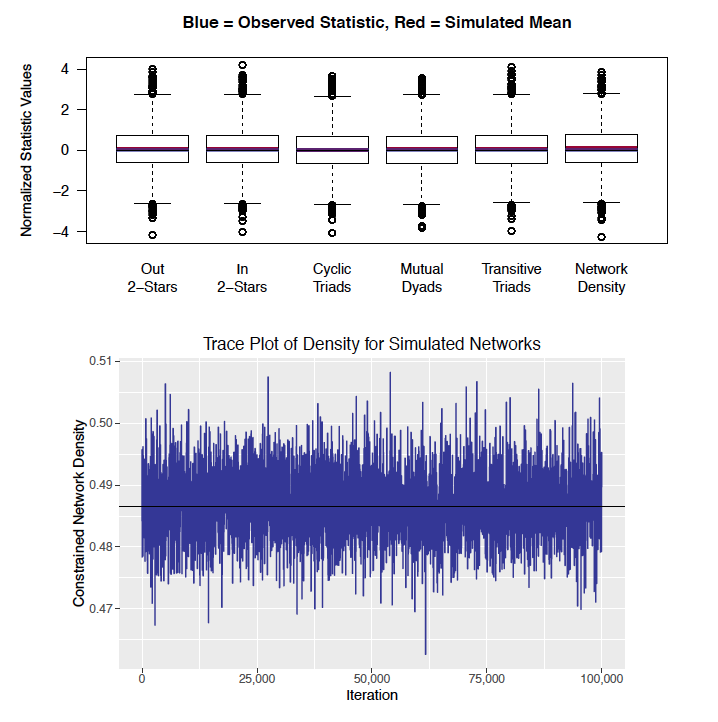}
\caption{\small U.S. migration network trace plot for the network density of the simulated networks from 100,000 simulations from Metropolis--Hastings. \label{fig:gof}}
\end{figure}

%In each plot, we compare the distribution of the network statistics generated from the 10000 simulated networks with the true observed network statistics of the migration network. Figure \ref{fig:gof} reveals that the distribution of network statistics from the simulated networks of each fitted model closely matches the observed statistic from the migration network suggesting good fit of each method.

%\newpage
% \begin{center}
% {\large\bf SUPPLEMENTARY MATERIAL}
% \end{center}
%
% \begin{description}
%
% \item[GERGM R-package:] R-package containing {\it GERGM.R} and {\it MH-Sampler.cpp} functions. These functions were used for all simulations and applications in the numerical study described in this article. (GNU zipped tar file)
%
% \item[U.S. Migration Data:] Data set used in the illustration of the GERGM method. (.RData file)
%
% \item[Financial Data:] Data set used in the illustration of the GERGM method. (.RData file)
%
% \item[Numerical Scripts:] The R scripts used to 1) perform the analysis and 2) generate all plots in Section 4. (GNU zipped tar file)
%
% \item[Additional Figures:] Figure S1 - goodness of fit plot for the international lending network for four network statistics that are not shown in Figure 2. Figure S2 - plot of $h_E$ and $h_{ITS}$ for the simulation study when $\theta_E = -5$. Figure S3 - plot of $\Delta(h_E, \theta_{\text{ITS}})$ for the simulation study when $\theta_E = -5$.
%
% \end{description}

%\newpage
%Bibliography
\newpage
\singlespacing

% \bibliographystyle{agsm}
% \bibliography{gergm}

\end{document}